# Gummy's Way Out - a Tangible Interactive Narrative with Food and the Diegetic Body

Saumya Gupta

University of California, Irvine, saumya@uci.edu

Theresa Jean Tanenbaum

University of California, Irvine, ttanen@uci.edu

There is growing interest in designing playful interactions with food, but food based tangible interactive narratives have received less attention. We introduce Gummy's Way Out, an interactive tangible narrative experience where interactors eat a gummy bear and help him find his way out of their bodies by eating various food items. By consuming different things, the interactor either helps or hinders the gummy bear's journey through an imagined 'diegetic' body that overlaps with their own. Interactors are endowed with the gummy bear's well-being and are also encouraged to reflect on how their actions can impact their 'lived' body. We present preliminary results of a user study and design considerations on how to design for the diegetic body in interactive food based narrative experiences. We recommend leveraging the sensory and emotional properties of food to create a visceral narrative experience.

CCS CONCEPTS • Human-centered computing~Interaction design

Additional Keywords and Phrases: **tangible interactive narratives, diegesis, human-food interaction, lived body, food, diegetic interfaces**

## 1 INTRODUCTION

The field of food and technology in HCI called Human Food Interaction (HFI) has been dominated by optimizing food practices and nutrition tracking [2]. In recent years, researchers have introduced the importance of studying food and play, to engage with the social, emotional, hedonic, sensory, and cultural dimensions of food [2,28,35,40–42]

However, the intersection between food and tangible interactive narratives has received much less attention. Food can evoke emotional, sensory, and nostalgic responses that make it a valuable storytelling component [2,30]. Moreover, the visceral quality of consuming food makes a case for exploring tangible narratives with food that can engage our bodies [27,37]. Similar to playful HFI, tangible interactive narratives with food can elicit new ways of experiencing and appreciating food [28], experiencing different cultures [1] and fictional worlds [12], encouraging mindful eating [28], and turning one's attention to their body [27], among others.

The few works that do exist within the domain of food based interactive narratives often position participants as observers who are situated outside the story world and can't impact the story [10,39,40,48]. We advocate for making people active participants such that are positioned inside the story world (internal roles) [15] and can impact the narrative world (ontological role) [15]. This is important because enacting and having stake in the story world rather than passively witnessing can help internalize the story, reflect on beliefs, broaden world views by seeing different perspectives , feel more responsible, and understand the character's goals [11,29,36].

We introduce our creation Gummy's Way Out (GWO) – a light-hearted interactive food-based narrative where the participant eats a sentient gummy bear who is eager to get out of the participant's body (narrated through audio). The gummy bear (called 'Gummy') asks the participant to consume various food items and perform bodily actions to help him on his journey out as he crosses various organs in the body. The participant's actions impact their narrative body in ways that either help or hinder Gummy's journey, endowing them with the well-being of Gummy and their body. This not only makes participants a part of the story world (internal role), but also gives them a stake in the story as what they eat impacts their body and subsequently Gummy (ontological role), moving their role much beyond passive observation. We aim to open the design possibilities in food based interactive narratives through GWO.

We use 2 design concepts in GWO – 'diegesis' and the 'lived' body. Diegesis is something that exists inside the narrative world, like a song playing on a radio instead of a background score [4]. In tangible interactive narratives, diegesis refer to elements that exist in the narrative world as well as the real world [9,14,17,24,26,38]. Diegesis can bridge the participant's reality and the story world [26], connect them to characters [13], and engage their body [24]. In GWO, the participant's body and the food are diegetic as they exist inside the story world and are a part of their reality. The diegetic body also gives participants stakes, responsibility, and a way to impact the narrative, moving their role beyond passive observers.

GWO leverages food and bodily actions to bring the participant's focus to their 'lived body' – through which they can experience the world, feel sensations, and emotions rather than just treating the body as an interface [27]. The lived body can help participants tap into various sensory experiences, which can be valuable in interactive narratives. We designed GWO with the 'lived' body to help people feel their actions in the story more viscerally and engage with the sensory and emotional dimensions of food and the body.

We explore how the diegetic lived body can enrich the experience of a food based interactive narrative and provide design takeaways based on a study we conducted with 19 participants who experienced GWO. The contributions of this paper are:

1. Describing the design of a food-based tangible interactive story that positions the participant in an internal-ontological role (not a passive role)
2. Providing recommendations to design for the 'diegetic' and 'lived' body in food-based interactive narratives
3. Providing design possibilities in the nascent field of food and interactive narratives

We do not aim to give generalizable recommendations, instead we give design takeaways for exploring the intersection of food and tangible interactive narratives further. This is a relevant contribution as to our knowledge no research has designed and analyzed an interactive food-based narrative where participants are more than passive observers. There is also not much work on how to design for the 'diegetic' and 'lived' body in tangible interactive narratives. These contributions can be useful for the tangible interactive narrative community, and the area of playful HFI.

## 2 BACKGROUND

The intersection of food and Interactive storytelling is a nascent area. We encourage exploring this space further by taking concepts from tangible interactions and storytelling such as 'diegesis' and by giving participants more active roles by putting them in 'internal-ontological' positions. Concepts from food and play design such as the 'lived' body that can further enrich the area. We describe these areas in more detail below.

### 2.1 Existing work in food and interactive narratives position participants as passive observers

In existing food-based narratives, most experiences position participants in passive roles, where they are observing the events of the story without being a part of the narrative. For example, in Gustacine, viewers eat popcorn that changes flavor while watching a movie - sweet cinnamon for joy and bitter mustard for grief, among others [22]. In an edible cinema experience in London, viewers eat chocolates while watching Charlie and the Chocolate Factory [44]. In another London restaurant, digital projections on food and crockery tell the story of the chef [39]. An exception to this norm of passive experiences is the Matter Hatter experience, an early exploration that turns dining into a narrative, challenge-based social experience where participants take different roles and dive into the world of Alice in Wonderland [3]. Gingerline, an interactive dining company, has also explored interactive narratives ranging from dinging and story performances to administering food in unique ways [45]. While these works start to investigate how foods from a narrative world can be integrated into a dining experience, they are highly contextualized in a restaurant setting. Moreover, they don't leverage the concepts of tangible interactive narratives described below that can enrich the space further. In GWO, we explore how to give participants more active roles, by engaging the 'diegetic' body.

## 2.2 Tangible Interactive Narratives

Tangible objects have long been used in interactive narratives to give participants a gateway to the story world. Harley et al. provide a literature review of interactive stories that use tangible objects [15]. Chu and Mazalek extend this framework to include interactive narratives with bodily interactions [5].

While Harley et al. identified multiple interesting themes, the one that we focus on here is the narrative position [15]. This is divided into two dichotomies which describe the position of the participant relative to the story [49]:

- Internal vs. External: Internal participants see themselves as situated within the world of the story while external participants are outside of it.
- Exploratory vs. Ontological: Exploratory participants observe and might reorder events of the story, but make no changes, while ontological participants may impact outcomes within the story world.

An example of an internal-ontological system is The Breathtaking Journey (a VR experience), where the player takes the (internal) role of the protagonist, a refugee trying to escape (Figure 1a)[24]. As the protagonist hides in a truck, the player's breath is tracked and if the sound level of the breath is above a certain threshold, the protagonist is captured. In this way, the player's breath (a bodily sensation) is used to impact the narrative (ontological role). On the contrary, Dagan's work positions the participant in an external-exploratory role [7]. Here the participant wears the coat of a character and rummages through the pockets to find different objects and hears the character's stories related to those objects. Participants are not a part of the story world (external role) and can only explore different segments of the story without impacting it (exploratory).

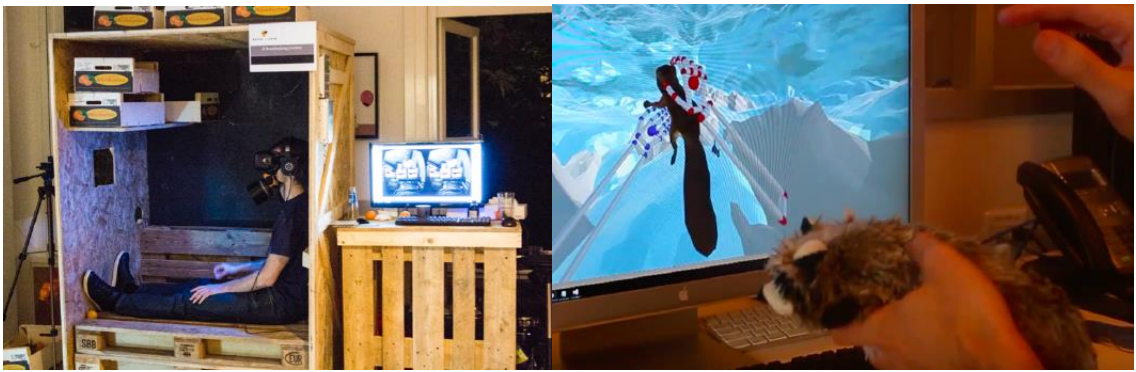

Figure 1: a) The Breathtaking Journey [22]                    b) The squirrel in [16]

Harley et al. pointed out that many interactive, tangible narratives place people in external-exploratory roles but only few place them in internal-ontological positions [15]. Internal roles allow people to enact rather than witness the story, which can help them integrate the character perspective in their own lives, and internalize and personalize story [11,29,36]. Ontological roles can give participants stakes in the story which can make them feel more responsible, invested [13,36], and help derive complex and different meanings [11]. Internal-ontological roles in interactive narratives are worth exploring, not just in food-based interactive narratives, but also more broadly in TEI. In GWO, we explore how to position the participant in an internal-ontological role.

Tangible objects have long been used in interactive narratives to give people a gateway to the story world. A commonly used concept in this field is 'diegesis'– elements that exist inside the narrative world [4]. For example, a song playing on the radio in a move is diegetic, but a background score is not. 'Diegetic objects' in tangible interactive narratives are objects that can exist in the narrative world as well as the real world [9,14,17,24,26,38]. These diegetic objects in the real world are not just mere representations of objects in the story world, rather they are the objects in the story world.

A classic example of a diegetic interactive narrative is genieBottles where participants open bottles and hear different genies come out to tell their stories [26] (Figure 2a). The bottles are diegetic as they exist in the participant's physical world and are homes to the genies in the story world. Another example is the Reading Glove where people interact with unique diegetic objects from the story to traverse through a non-linear thriller narrative about a spy [38]. Participants expressed how they could identify with the main character by holding and interacting with diegetic objects that the character interacted with (Figure 2b). Shiva's Rangoli uses a diegetic Rangoli interface to help participants take an internal role of God Shiva and create Rangolis on his behalf to express his emotions [14]. Diegetic objects in virtual reality narratives are also upcoming [16–18]. In [17] the people interact with a tangible, diegetic squirrel which they can hold in reality and see in VR (Figure 1b). They can feel the squirrel's heart rate through haptic interactions and slow it down by stroking the squirrel. Here too participants take an internal-ontological role as they play a role in the squirrel's world and can impact the character.

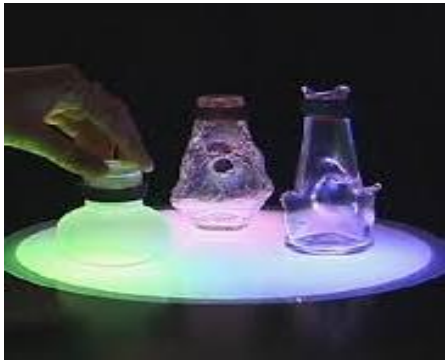 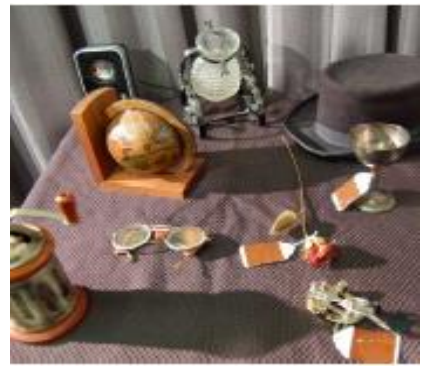

Figure 2: a) genieBottles [24]                                    b) The Reading Glove [36]

Beyond diegetic 'objects', a few examples explore the diegetic 'body' as well – where the participant's body is also a part of the narrative. Tangible Comics is a full-body interactive performance where the participant plays the role of a female egg in the reproductive system and uses diegetic objects and gestures such as jumping and waving [32]. This work shows the potential of using full-body interactions. A Breathtaking Journey described above is a rare example that shows how the participant's diegetic breath can be used to place them in an internal and ontological role as their breath impacts the narrative [24], a concept we use in GWO as well. The authors of [43] re-create Flatland, a 2D world with only sound and touch as the prominent senses. The audience members step into a dark interactive theatrical experience where they cannot see but only hear snippets from the story world by interacting with diegetic pieces like pipes and windows. Although these works position the body in a diegetic role, none of them focus on the design and analysis of the diegetic body.

All these works above show how diegesis can give people a gateway to the story world, feel a part of the narrative, engage their bodies performatively, and even connect with characters in the story. Diegesis can be particularly useful in food-based interactive narratives, where diegetic food can help participants enter the story world and possibly impact the narrative. As per our knowledge, GWO is the first project that discusses the design and analysis of diegetic food and the diegetic body in an interactive narrative experience.

### 2.3   Food and Play Design

Besides tangible narratives, the area of food and play also has design concepts that can further enrich the intersection of food and interactive storytelling. We focus on the concept of the 'lived' body. While the existing work with HFI [2] and body experiences[8] is vast, for GWO, we focus on the intersection of food, body, and play.

Mueller et al., inspired by Merleau-Ponty's phenomenology, use the German terms 'Korper' (corpse) and 'Leib' (lived body) for bodily play [27]. They give examples of designing for the lived body through which people can experience the world, feel sensations, and emotions rather than just treating the body as an interface. For example, placing a big button that needs to be pressed with both hands, at a height above one's head level, would require a person to raise their arms. This would put people in a 'winner-pose' which is associated with positive experiences, which brings the Leib into perspective (how the person feels when they perform the action). They further discuss that a person's focus can be brought to their lived body through perception and localized sensations (touch, pain, proprioception, kinesthetic sensation, and temperature perception), as these sensations enable people to experience their bodies as 'theirs' (Leib). Tanenbaum and Bizzocchi described the pleasures of using the body through the game Rock Band where players act like rock band musicians [37]. Players are incentivized to engage their Lieb by moving their bodies into a stereotypical guitar pose to activate special abilities in the game.

Svanæs discuss Merleau-Ponty's description of abstract vs. concrete movements, another concept that can bring people's focus to their lived bodies [34]. Movements made naturally as a part of daily life actions like walking are 'concrete'. However, if a person is intentionally asked to move their left foot in front of the right, outside the context of walking, the action becomes 'abstract', and forces people out of their habitual behaviors, bringing focus to their lived bodies.

Simialry, Höök et al. used Merleau-Ponty's phenomenology to describe Somesthetic design - turning inwards through experiences that encourage slowness, delicate touch, and care to bring attention to the 'pulsating, live, felt' body [19]. The authors describe the design of curtains with lights that fade out in sync with one's breath, to encourage turning inwards.

Many social experiences with food apply the concept of the 'lived' body. TastyBeats is a playful fountain installation where players get a customized cocktail based on their heart rate data from physical activity (Figure 3a) [21]. In Mad-Mixologist, players need to mix ingredients for a potion together, but their visions are swapped through their headsets [46] (Figure 3b). Arm-a-dine is a two-player experience where prosthetic arms strapped on the player's waists, feed the players based on their facial expressions (Figure 3c) [28]. In the Guts Game, players eat ingestible temperature sensors and compete with each other by changing their body temperature [25]. In all these examples, playing with food brings one's focus to the lived body. In GWO, we leverage the concept of the 'lived' body so participants can feel the sensory and emotional impact of their actions in the story.

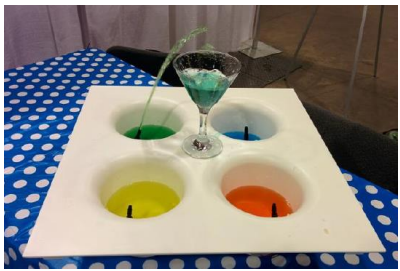 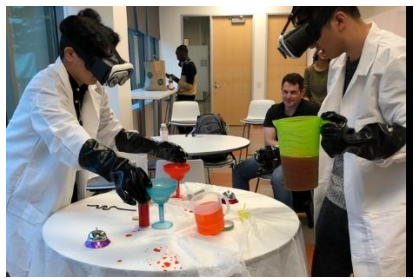 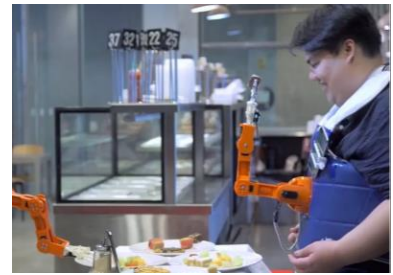

Figure 3: a) TastyBeats [19]    b) Mad Mixologist [44]    c) Arm-a-dine [26]

There are a few audio and food-based playful experiences that have inspired GWO. Chewing Jockey is a technical investigation where users hear different sounds as they chew and perceive a changed flavor profile [23]. The authors mention that users were amused when they chewed on jellybeans that made screaming noises. We used this finding in GWO and took the idea further through a narrative (more in design section). In Lickestra, users make different instrumental sounds by licking ice cream [47]. The authors note that roaring sounds can transition people into a fantasy world, another promising finding for interactive narratives. All these examples show the potential of interactive food experiences, but none of them involve storytelling. GWO leverages this potential and enables participants to interact with a food-based story as they hear the story through audio.

## 3 DESIGN

Gummy's Way Out is a light-hearted humorous story of a sentient and sassy gummy bear (called Gummy) who gets stuck in the participant's body and tries to find his way out with the participant's help. This is a seated experience where participants hear Gummy's dialogues through audio and help or hinder Gummy by consuming various food items (Figure 4).

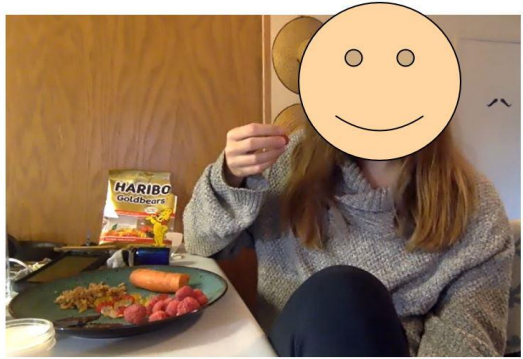

Figure 4: Interacting with GWO

Gummy travels in the participant's body from organ to organ, giving visceral descriptions of what he sees, the hurdles he faces, and how the participant could help him. As the participant consumes different, he describes how the item is impacting the participant's body (narratively), and subsequently Gummy (Figure 5). Gummy's health depends on the body's health, which endows the participant with not only Gummy's well-being, but also their own body's.

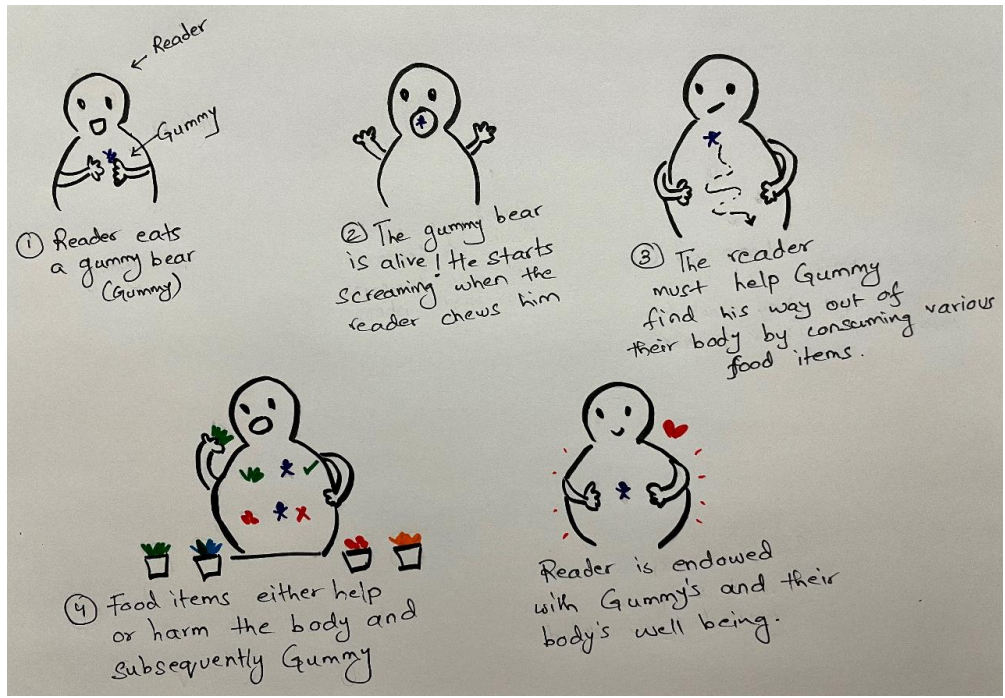

Figure 5: Illustration of GWO interaction

The participant's body is positioned in a diegetic role as the story happens inside their real body, which is also a part of the narrative world [4]. Gummy constantly calls attention to how the participant's actions feel and impact their body, which calls attention to their 'lived' body.

This design also puts participants in an internal-ontological role as their bodies are a part of the story and they can impact Gummy's journey [15]. While they are encouraged to think about food's nutritional components, this experience is not meant to be educational, and is more about focusing on the lived body and building a relationship with Gummy.

The food spread consists of - two different servings of healthy foods like fruits and vegetables, an unhealthy oily food, an unhealthy sugary food, a drink that one finds relaxing and would have before bed, a drink that one finds energizing and would have in the morning, water, and a few gummy bears (Figure 6). Within these broad categories, participants are free to choose and personalize food items they want to consume for the story experience (see study section for more details).

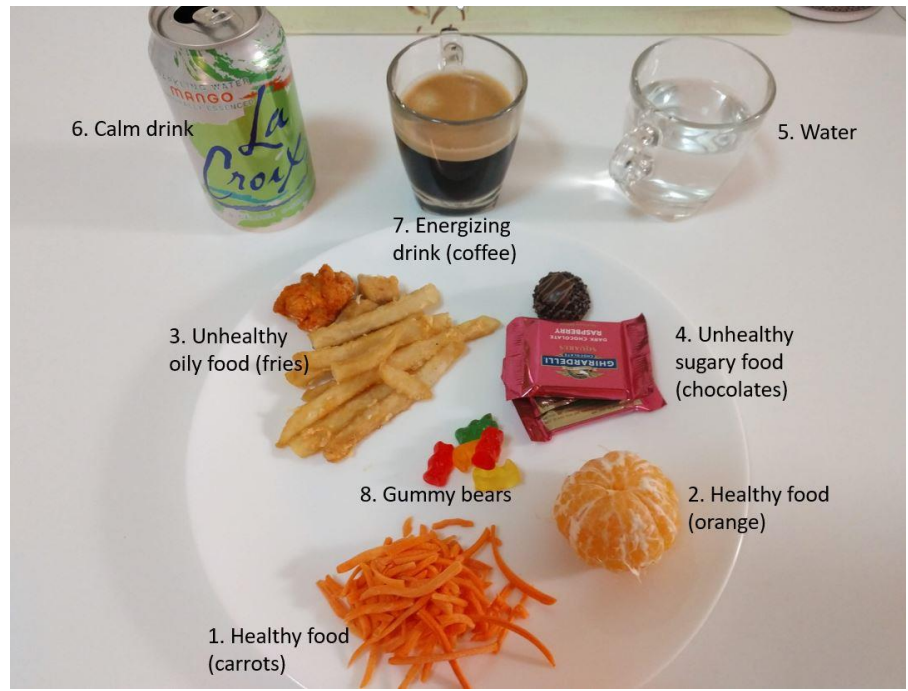

Figure 6: A sample of the food spread

### 3.1 Experience Walkthrough

The story starts when the participant eats a gummy bear. As the participant chews the gummy bear, they hear him (Gummy) cry out for help, begging the participant to stop (inspired by [23]). After being swallowed, he says he is in a dark pit – the stomach. From here on, Gummy asks the participant to help him find his way out of the participant's body. This scene hooks the participant to feel guilty and obliged to help Gummy.

Gummy describes what he sees to help participants visualize the story. In the stomach, he says he is in a dark pit which is accompanied by a reverb in his voice making it sound like he is in a big chamber. He describes the 'fleshy stomach walls churning food into pieces', accompanied by sounds of gurgling and churning. He sees a 'river of acid' that he must cross and asks the participant to start eating so he can make a boat out of the food. They are encouraged to explore how different food items impact the narrative. Healthy food helps Gummy build a boat and keep the acid in check and unhealthy food does not. For example, when the participant eats oil unhealthy food, Gummy says:

> "Ouch! I am slipping on all the oil you just ate. Your fleshy stomach walls seem to be working too
> hard, it's scary!"

Throughout the story, Gummy acknowledges how the participant's action feels, and how it impacts the participant's body and Gummy. If they eat more unhealthy food in the stomach, the acid rises, and Gummy expresses that he is burning, and that the participant needs to curb the acid.

> "Uh oh, what did you just eat? All that oil! it is making this acidic river rise!  (Swirling acid sounds).
> Too much acid…this river is spinning out of control, (screams) Hurry! drink something to tame this
> acidic river, your belly walls are on fire! (Acid swirls, stomach gurgles) aaaah it hurts!"

 If the participant continues to eat unhealthy food, the situation worsens, and Gummy complains about his limbs burning and the stomach walls getting bruised because of the acid. These actions not only hurt Gummy, but also the participant's narrative body. Through such worsening consequences, they are nudged to curb the acidity. After doing so, Gummy describes how he narrowly escaped and that he can see the stomach walls cleansing. He urges the participant to drink something calming, so they can both relax after the fiasco. Gummy and the participant enjoy a slower, calmer moment, encouraging them to keep sipping the relaxing drink (something they drink before going to bed):

> "Ahhh (long sigh), that feels nice, very nice. Every corner of your stomach is softening. I'm 'finally' going to take a few moments to relax by this tranquil river, you should do the same! Keep sipping".

Participants can also choose to drink caffeine instead, where a grumpy Gummy would curse the participant and move ahead.

Next, Gummy sees the way out of the stomach and asks for a belly rub to help push him along. This is the first body action that participants perform which is different from food consumption. As they rub their belly, Gummy says he feels the warmth and the massage. Participants could have chosen not to help Gummy, where he would have expressed distrust and waddled his way out. Either way, Gummy lands in the small intestine where he gets stuck in the villi. Participants must eat another gummy bear called Ana, who finds Gummy and helps him out. They both cross a few more hurdles with the participant's help and end up getting absorbed from the intestine into the bloodstream (like most food). Gummy loses Ana during this scene and his mission from there on is to find her and get out.

Gummy lands up in the spleen to find Ana, where he encounters white blood cells (WBCs). The white blood cells see him as an outsider and start attacking him. There are alarm sounds ringing 'intruder alert' and laser shooting sounds, as Gummy runs, and shouts. This is the only scene where the participant's actions hurt their body but help Gummy. If they choose to help their white blood cells by eating healthy food, the WBCs get stronger, Gummy gets shot, questions the participant's loyalty, and after three attempts, he dies. If the participant eats unhealthy food, their WBCs slow down giving Gummy a narrow escape.

Next Gummy enters the bowels thinking he might find Ana there. But meets a massive pile of 'hard brown rocks'. Participants must eat food with fiber to help Gummy, if not, the pile of rocks gets bigger, risking an avalanche. Gummy asks for another belly rub to proceed from the bowels, and then claims to see a light – the way out of the body. However, he goes back into the bloodstream as he decides not to leave without Ana.

A dejected and helpless Gummy lets the bloodstream take him, as he thinks about Ana. There is a thumping heartbeat sound in the background and Gummy says he is drifting into the participant's heart! He asks them to slow down their heart rate as it beats too fast for him to waddle in. As participants take deep breaths or drink their calming drink, they hear the heartbeat sound slowdown in the background. They can choose not to help Gummy, where he would get a bumpy ride and guilt the participant. If they help Gummy, he expresses his gratitude by massaging their heart.

Either way, he still cannot find Ana and starts to imagine the worst-case scenarios, thinking his way into a panic attack. He gasps and urges the participant to take deep breaths and send in oxygen. Participants also hear the heartbeat thumping increase in pace again. If they take deep breaths, they hear the heart rate slow down, and Gummy expresses his recovery as he feels less claustrophobic (inspired by [24]). Soon after, Ana pops into the heart from her own quest of searching for Gummy. They reunite and exchange stories, they both thank the participant, and the story ends with them traveling to the bowel to find their way out. At this crucial end scene, if the participant decides not to take deep breaths during the panic attack, Ana still finds Gummy, but only to find him dying. This is where the experience ends.

While participants can go against Gummy, they are nudged into taking the desired action through worsening consequences like Gummy losing his limbs in the acid attack and spleen or questioning the participant's loyalty. Gummy's physical and emotional condition, the condition of the body itself, Gummy's trust in the participant, all depend on the participant's actions. In this way, their choices impact Gummy's attitude and their relationship with him and their bodies, but not the story's plot, maintaining a linear story while still giving them ontological choices. GWO went through two design iterations before reaching this described state.

## 4 IMPLEMENTATION

Although the study was conducted online due to the Covid-19 pandemic (more details in study), we implement a version before. The food and drink items were connected to the computer through a Makey Makey [33]. The Makey Makey board was clipped to different spoons and forks made of steel, that were mapped to different food items. A small copper tape was placed on the rims of the glasses. Participants sat with a glove on their dominant hand, so the circuit was not completed when their hand touched the utensils but would be completed when their lips touched the utensil, which would trigger the audio clip associated with that item based on the story segment (Figure 7).

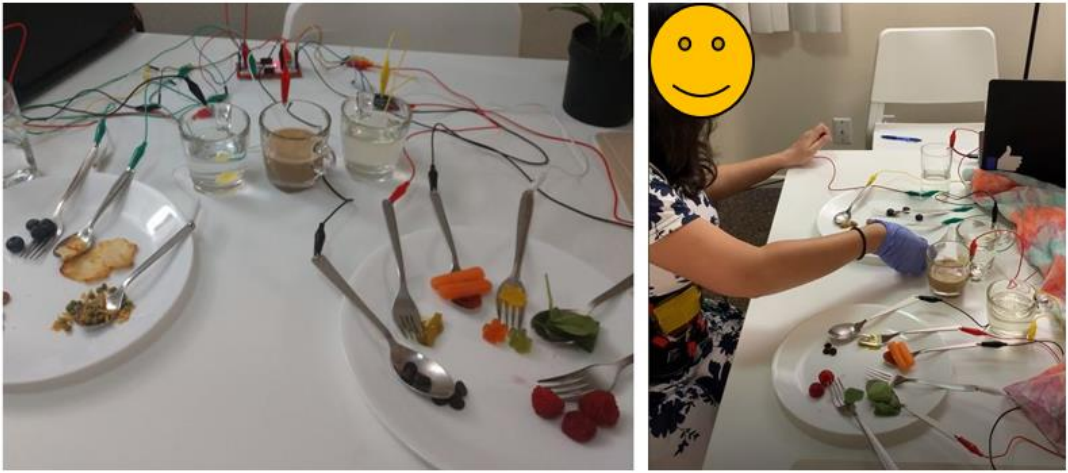

Figure 7: Utensils connected to the Makey Makey and a participant interacting

Participants also wore a belt that contained mini vibration motors and Adafruit's Flora [20] (Figure 8). The Flora was plugged into the laptop using a long USB cable, which was convenient as this was a seated experience (Figure 7). In this design with the belt, the vibrations signified Gummy's movement inside the stomach and intestines. The Python script handled interactions with the Makey Makey, decided which audio clip to play, and sent commands to the Flora for the vibration motors. As detailed in the next section, we ran this study virtually in a wizard-of-oz style due to the Covid-19 pandemic and did not end up using the belt. We added deep breaths and belly rubs in the design to bring more focus to the body, in the absence of the vibration motors.

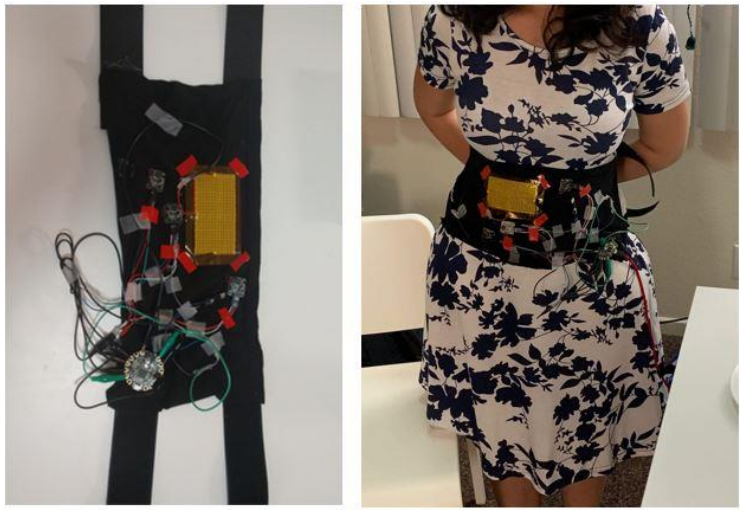

Figure 8: The belt with the vibration motors and Flora

The mechanism for belly rubs were implemented through an alligator clip on the belt around the waist connected to the Makey Makey. Once touched by the non-gloved non-dominant hand, the circuit was completed. Similarly, for a low-tech solution, we asked the participant to touch their chest while taking deep breaths. While this gives a faster and cheaper way to technically input the act of deep breathing, it also adds to the design of the lived body. Touching one's chest is more performative as it allows one to feel the movement of the body while taking deep breaths.

## 5 STUDY

The goals of this study were – 1) Analyze how participants interacted with the diegetic 'lived' body while taking internal-ontological roles in a food based interactive story 2) Understand what the diegetic, lived body added to the interactive narrative experience. The story we used (described in section 3) was not meant to be 'educational' as the line between fact and fiction was blurred, it was meant to answer the goals described above.

Table 1: Participant demographic data

| Participant ID | Age | Gender | Occupation |
|---|---|---|---|
| P01 | 35-44 | Female | UX Designer |
| P02 | 35-44 | Male | Graduate Student |
| P03 | 25-34 | Male | Graduate Student |
| P04 | 25-34 | Female | Graduate Student |
| P05 | 25-34 | Male | Graduate Student |
| P06 | 25-34 | Female | UX Designer |
| P07 | 18-24 | Female | Graduate Student |
| P08 | 25-34 | Female | Graduate Student |
| P09 | 25-34 | Female | UX Researcher |
| P10 | 25-34 | Male | Graduate Student |
| P11 | 18-24 | Female | Undergraduate Student |
| P12 | 25-34 | Female | Graduate Student |
| P13 | 25-34 | Male | Graduate Student |
| P14 | 35-44 | Female | Graduate Student |
| P15 | 25-34 | Female | Graduate Student |
| P16 | 25-34 | Male | Graduate Student |
| P17 | 25-34 | Female | Graduate Student |
| P18 | 35-44 | Male | Graduate Student |
| P19 | 25-34 | Female | Graduate Student |

We conducted an unpaid study in December 2020 with 19 participants (13 female, 6 male), recruited through convenience sampling, ages between 21-40, all based in the US, and worked as researchers, designers, and graduate students (Table 1). All participants had engaged with some form of interactive narratives through museums, choose your own adventure story books, video games, and RPGs. Since the study was conducted during the Covid-19 pandemic, all sessions were held virtually on Zoom.

We instructed participants to plate items from the food categories - two different servings of healthy foods like fruits and vegetables, an unhealthy oily food, an unhealthy sugary food, a drink that one finds relaxing and would have before bed, a drink that one finds energizing and would have in the morning, water, and a few gummy bears (Figure 6). All items were based on the participant's preferences, and adjustments were made based on dietary needs. Figure 9 shows a few of the participant's plates.

We conducted the study sessions in a 'wizard-of-oz' style study where the researcher kept track of what the participant ate during the narrative and shared their computer's audio through Zoom. Participants did not have access to the belt and hence the vibrations were not included. Each session lasted two hours. The researcher briefed participants that they would be interacting with a light-hearted story by consuming what they had plated and instructed them to think aloud. The story experience lasted between 30-40 minutes, and the researcher took observation and think-aloud notes. After the experience, the researcher conducted an hour long semi-structured interview, asking them about their experience – an overview of what they found engaging, what they chose to eat and why, how they felt towards Gummy and why, how they experienced their bodies and what brought focus to their bodies, and what these different design elements did for their relationship with Gummy, if anything at all.

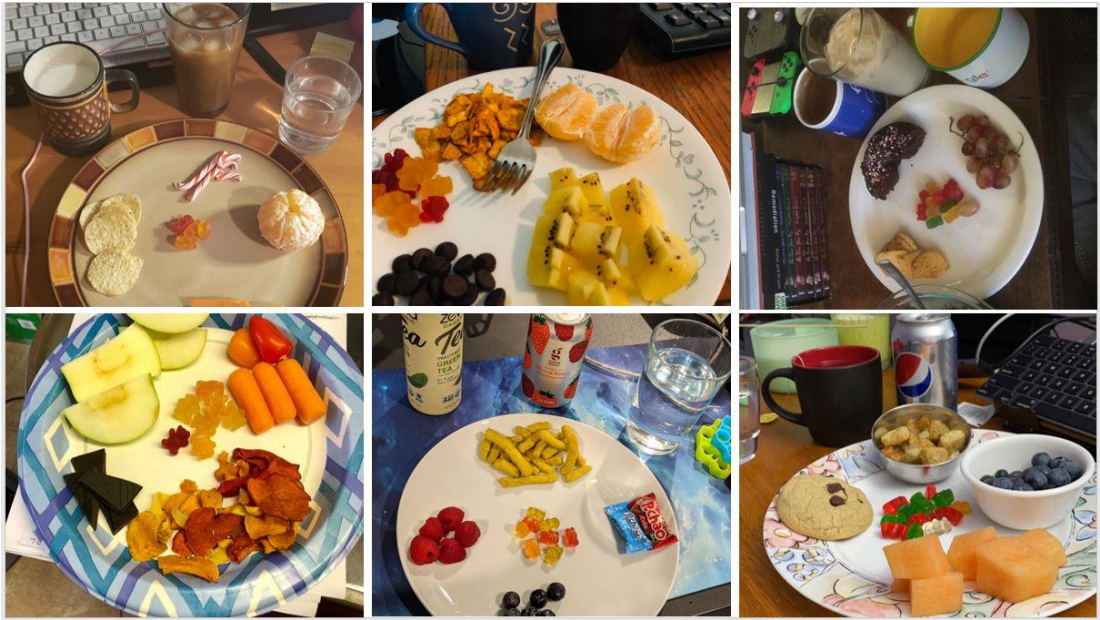

Figure 9: Plate setups of different participants

Despite the virtual study, we took steps to ensure the story experience was realistic. For example, participants were asked to use headphones, so the sound felt like it was coming from within them, they consumed real food, and they performed the bodily actions in the narrative.

We transcribed the interview data and then conducted an open coding analysis and then a thematic analysis of the transcripts and observation data based on the research questions. We then categorized the open codes, identified patterns, and created themes [6,31]. We went through each theme and how they connected with each other to brainstorm design takeaways. Major themes and takeaways are described in the next sections. These design takeaways are not generalizable but provide insights to create interactive narratives with diegetic food items and the diegetic body.

## 6 FINDINGS

We describe what helped participants feel their bodies were diegetic, the consequence of positioning the body in a diegetic role, and what hindered this process. We also discuss how participants engaged with Gummy by aligning with his goals and feeling responsible for him since he was in their body.

Participants felt their bodies were a part of the story world due to actions of deep breaths and belly rubs, and by focusing on what was happening inside the body especially when Gummy did something strange inside. This made them buy into the narrative consequences, added personal stakes in the story, and made them feel more responsible for Gummy as he was in THEIR body. However, a few things hindered participants from buying into the narrative, specifically when they could not recognize Gummy's form, and when there were anatomical inconsistencies. This break in diegesis made them feel distracted.

Since the participant's body was a part of the story world, they felt they had personal stakes. The participant's and Gummy's goals of helping the body, largely aligned which helped them connect with Gummy. Almost all participants felt guilty for chewing Gummy and putting him in their bodies, which made them feel responsible for him throughout. Participants also reported enjoying moments of calmness with Gummy especially with the deep breaths. The sub-sections describe these findings in detail.

### 6.1 Positioning the Participant's Body in a Diegetic and Lived Role

Almost all participants reported that some parts of the story felt like it was happening inside their body, but a few parts were playing out in their heads (spleen). They stated that the unexpected bodily actions of belly rubs and deep breaths reminded them that the story was happening inside their bodies. 8 mentioned that belly rubs were particularly useful for making them feel like they were physically interacting with Gummy, especially when Gummy acknowledged their touch and responded with gratitude. This helped them feel like Gummy was inside THEIR body and that they helped Gummy on his journey. These people also reported feeling that their body was 'living', which brought focus to the gurgling / churning happening inside and encouraged them to take care of their stomach.

> P7: "I think the times where I really felt like a connection when he asked me to rub my tummy. You pat your stomach and then you sense that he's there because he responds and says thank you."

> P9: "I think the audio feedback was that the gummy bear was comforted. And that just kept me going and I felt like it was actually translating through my body skin flesh to my stomach. So I felt very aware of my body at that time."

Almost all participants mentioned that the action of taking deep breaths helped them feel their body was a part of the story world as the action made sense logically in the narrative – whether deep breaths were taken to give red blood cells oxygen, for alleviating the panic attack, or for feeling calm. Since deep breaths served the same purposes in the story as in real life, participants were able to buy into the fact that the story was happening inside their bodies. This sentiment was not shared with belly rubs as many reported that they did not associate belly rubs with anything in real life.

> P6: "Whenever it asked me to breathe for whatever reason, it felt justified. When it said it wanted oxygen, then it connected to the fact that hemoglobin needs oxygen, when you are stressed you need to deep breathe to calm down."

Participants reported that the consumption of various food items throughout the experience helped them feel their actions viscerally. 3 mentioned they viscerally felt the oily and acidic food as they caused the stomach acid to rise in the story.

> P13: "I was eating different things, I experienced differences at each point in the story, not just based on the name of the food, but like the action of the food that I was eating"

The impact on the participant's body in the narrative (referred to as narrative body) helped them feel their body was a part of the story world. 5 people elaborated how any good or bad impacts on the narrative body brought focus to their bodies. For example, P11 mentioned how the cookie causing the stomach acid to rise was memorable for her. P17 felt her stomach tighten when she ate something acidic as she felt the acidity in her mouth and heard Gummy complain about the rising acid in her stomach. Similarly, a few felt grossed out by the oil in their mouths when Gummy slipped on oil. 7 mentioned that the heartbeat thumping felt like it was their own because they could impact it through deep breaths.

> P11: "when I ate like the cookie, the walls of my stomach were like, freaking out and the acid. Felt I should help the gummy and, oh, shoot, I shouldn't have messed with my stomach like that."

> P17: "I felt a tightness in my stomach. it's because my body was part of the story, my stomach was getting upset, I could feel the acidity (in my mouth), I could feel the heartbeat, it brought realism. The gummy bear was talking about being in there, and then the addition of more sugar and eating food I think it made it tight."

5 people reported that Gummy brought their attention to their bodies because of what he was doing inside their bodies. They were anxious of Gummy coming close to a gentle and vital organ like the heart, and the heartbeat sound added realism. One describes Gummy as 'something slithering in my heart'. 2 felt similarly during the spleen scene, where one felt queasy because Gummy let germs escape from the spleen. 3 people commented they felt uncomfortable when Gummy was trying to exit their body because they thought of going to the bathroom to let Gummy out.

> P5: "She says I'm gonna massage the inside of your heart for you! That sounds bizarre, like the fact that I have to pretend that a piece of food has arrived in my heart in the first place is terrifying. But then, but then she says let me massage your heart for you. I don't want to even imagine the sensation of something inside my body, like rubbing up."

> P1: "There was something very real about eating the gummy bear and then hearing it talk to you. I did have these moments of these brief twitches where I had this weird feeling of trying to imagine an actual sentient creature running around inside my body and yelling at me. And that was that did have like a certain visceral impact."

While the novelty of interactions, each interaction (belly rubs, deep breaths, food consumption) was repeated multiple times throughout the story so participant's experience was not just limited to a one-time interaction.

### 6.2 Consequences of positioning the body in a diegetic lived role

What did the diegetic lived body add to the narrative experience? 4 participants discussed how their body being a part of the narrative experience helped them buy into the story more as the body gave a sense of realism. They reported that the taste, consequences inside their bodies, and the story world inside all felt real.

6 participants said that their personal stakes in the story were higher as their bodies were a part of the narrative, increasing the gravity of their choices in the story. They felt their choice carried more weight as it affected their bodies inside and outside the narrative.

> P13: "Usually in a game when you take an action, it doesn't impact you in (reality) anyway. But here you know when you ingest something, it's really being ingested. It's not just gummy being affected, it's me too. The fries and carrots exist in my body. So, in screwing over gummy, I am physically screwing over myself."

People also acted in ways to help their bodies, since they had personal stakes. 7 said they took actions in the story such that they would help and not hurt their bodies in the narrative. Two mentioned how they felt bad for their stomach after the acid scene and wanted to eat healthy after.

> P5: "I felt bad eating sugary food and then experiencing how that was affecting these characters in the context of my body. And then of course, thinking beyond just this moment - like, Oh, my God, sugar consumption, in general, must be creating this atmosphere and, and it's weakening the white blood cells."

8 people mentioned they felt more responsible for Gummy as he was in their body, and hence was their responsibility. They felt whatever happened to him was because of their actions and their body's response. They also felt guilty of chewing him as well.

### 6.3 Breaks in diegesis

A few design concepts led to participants questioning the diegetic nature of their bodies and Gummy. 6 people questioned how Gummy was alive even after they chewed him. 2 also questioned Gummy's form as they could not imagine how he was traveling through their bodies in one piece. 3 pointed out how anatomical inconsistencies caused distractions – Gummy going from bowels to blood and how food (Gummy) was in their heart. 5 also pointed out they did not feel what was happening to their narrative body in a few scenes – Gummy trying to exit, and their heart rate increasing when Gummy was getting a panic attack. A few people also questioned why eating healthy food gave Gummy a boat. These breaks in diegesis caused distractions as people jolted out of the story.

> P18: "I destroyed the gummy immediately by chewing him…But then I was like, no, that's not what actually happened. He got mangled around like he was in the laundry machine. And so he was never actually torn apart…. I had to make that decision because that (Gummy being destroyed) doesn't work for how I want this story to unfold."

A few people mentioned different ways in which they perceived limited agency. 2 people felt limited agency as they felt they could not control their body's involuntary responses feeling helpless to help Gummy. However, they felt more in control while performing deep breaths, belly rubs, and chewing as these were actions they could control.

## 6.4 Connecting with Gummy

5 participants talked about how their overall goals of helping Gummy and helping their bodies aligned with Gummy's goals of getting out of their bodies. They felt they would have to go against their own bodies to harm Gummy and mentioned they liked that Gummy cared for their body's well-being too.

The most successful design element that helped participants feel responsible for Gummy was chewing him in the beginning. 12 people felt like they were hurting Gummy as he was screaming. They felt guilty of eating something that was alive especially when he begged them to stop, and they still kept chewing him. The chewing scene set them up for pledging their allegiance to Gummy throughout the experience, as they did not want to defy him.

> P09: "It's almost like when you're chewing it, you are chomping, plus that audio experience gives a very dramatic effect. feel like a monster eating a tiny human."

> P05: "I am emotionally invested in her as a character, she was really sympathetic early on, so, I was reluctant to defy her. When the first gummy bear revealed themselves to have consciousness and the ability to talk to me, I felt immediately guilty about the fact that I was hurting them by chewing."

> P15: "I don't think it ever crossed my mind to mess with gummy's journey…. I think I was just kind of invested in gummies journey beginning …if the beginning had not been quite as empathetic, I may have felt like why should I care?"

The belly rubs, deep breaths, and calming drink were designed to help participants share a calming moment with Gummy. 4 people found belly rubs comforting, and Gummy's reaffirmation further enhanced the feeling. However, 8 others did not share this sentiment as they found belly rubs to be an unusual action, something they did not do in real life. 2 questioned why Gummy acknowledged the warmth from the belly rubs as it did not help him in any way.

The calming drink was successful in making most people feel calm. 10 people felt calm because of their body's physiological response, but 4 others reported that the link between the drink and calmness made sense logically (cognitively), and they felt relief because of how Gummy felt. Personalizing the calming drink for every participant helped enhance the feeling. The deep breaths followed a similar pattern where people found it calming physically, or because of gummy.

> P16: "There's like the first moment I drink the calming drink and she said, oh, I feel relaxed. I feel relaxed too. Maybe that's the most comforting moment during the entire experience…. the calming down really syncs with my body…Like if this is how my body feels at that moment, if the Gummy in the story feels the same way as my body feels, then yeah, I feel it too."

> P15: "It was very calming. I found myself closing my eyes when I need to take deep breaths and focusing on my breath. And I feel like that was really engaging for how my body was feeling throughout the story…. I think it was calming because of the deep breaths initially, but then, it was like a validation that Gummy also found it comforting…. it helped center my body in the story as well, because I saw or heard the immediate ways that like, doing something with my body had an effect in the storyline and emotions of the character that I was following."

## 7 DESIGN TAKEAWAYS

GWO showed how the diegetic body can add personal stakes, increase the gravity of choice, help participants feel actions viscerally so there is more realism, endow participants with responsibility, produce affective responses, and feel connected with the character. Many people described the experience as real and visceral. We describe how to design for the diegetic lived body in a food-based interactive narrative to achieve the above.

We observed that to make participants feel the above, they must buy into the fact that it is THEIR body that is a part of the narrative and feel their actions viscerally. We give three design recommendations– 1) creating body tethers that are design hooks that remind participants THEIR body is in the story world and is diegetic, 2) leveraging the emotional and sensorial aspects of food and tying them into the narrative consequence and 3) leveraging the diegetic body for character connection

### 7.1 Body Tethers

In tangible interactive narratives, the diegetic interface gives participants a tangible tether to the story world. For example, in genieBottles [26], the bottles were a gateway to the story, reminding participants how the story world was connected to their reality. However, when the diegetic body is a gateway to the story world, participants may forget that the narrative is referring to THEIR body, as they are not always aware of their bodies in space like they are aware of objects in front of them. For example, in GWO some participants mentioned that they imagined parts of the story in their head, rather than in their bodies. We recommend creating body tethers as hooks that remind participants that their body is a part of the narrative world. This can then add personal stakes, gravity of choice, and help participants feel their actions more viscerally.

Body tethers can be created by asking participants to perform unfamiliar and out-of-the-ordinary actions that bring focus to one's body such as – taking deep breaths or belly rubs. In GWO participants mentioned such actions brought focus to their bodies. Svanæs describe that concrete movements are everyday movements like walking and abstract movements are made outside normal context, bringing people out of habitual behavior such as asking a person to move the left foot in front of the right [34]. We recommend that performing abstract movements outside normal context can act as tethers for the diegetic body.

Just performing an abstract action in the narrative is not a complete body tether. We recommend that the story should also acknowledge how the action feels and impacts the narrative. For example, Gummy responded to deep breaths by saying they felt calming (how the action feels) and helped lower the heartrate (impact of the action). Acknowledging how the action feels can help people experience their body as theirs (lived body) as it brings attention to localized sensations [27]. However, if the action does not have a narrative impact, people may not notice the acknowledgement at all. For example, not many people noticed the warmth of the belly rubs as it had no narrative consequences. We elaborate on this topic further in the next section.

Tethers can also be augmented by bringing attention to the physiological response of an action - such as hearing a change in heartbeat after taking deep breaths [24]. They can also be facilitated by a character that shows agency over the participant's body with conflicting interests. Participants mentioned they had a visceral reaction to Gummy moving towards their heart, and when Gummy asked them to eat junk food for him to escape the WBCs, creating a moment of conflict for the participants.

Designing with the diegetic body can be complicated as the body is very close to the real world, making suspension of disbelief more challenging. Participants expected Gummy to follow the rules of their body as he was stepping into their reality, not the other way around. Since the story was grounded in the participant's reality, they got distracted when Gummy did anything anatomically inconsistent like traveling from the bowels to the heart. While designing with the diegetic body, it is important to keep in mind the fragility of suspension of disbelief, especially if the story is grounded in the participant's reality. The suspension of disbelief is harder to achieve when one's body is the interface, compared to physical objects [14,26,38].

Some participants were confused with Gummy's changing form throughout the story. Some participants felt like chewing Gummy broke him, which meant he was dead. This broke the participant's recognition of Gummy as a character. We recommend considering how an action may change the form of the diegetic interface and what participants might conclude from that.

Lastly, the diegetic body can create immersive but also uncomfortable experiences as they may bring attention to a participant's health problems. While it is important to design empathetically and avoid negative experiences, designers can also provide a note of caution beforehand.

### 7.2 Leverage the emotional and sensorial aspects of food and tie them into the narrative consequence

In GWO we explored how food based interactive narratives can help participants feel their actions viscerally, increase their buy in, and add stakes to impact the story world (ontological roles).

We recommend leveraging the sensorial, and emotional (affective) properties of food to engage the lived body and achieve the above. While existing work [27,37] describe how to engage one's body in play, they don't explore how food can add to the experience, and food-based experiences often position food only as a playful object [21,46] , they don't leverage the sensorial and emotional qualities of food.

In GWO, participants found certain parts of the narrative more visceral, engaged the lived body more, and added more stakes in the story than others. For example, a few participants mentioned they had a visceral reaction to Gummy slipping on oil after they consumed oily food. And they felt calm while taking deep breaths and drinking their 'calm' beverage. However, they had a tough time buying into why healthy food gave Gummy a boat in the acid river. Why was the oil and calm beverage example more visceral and believable than the boat one? Our analysis revealed two possible reasons:

1) The oil and calm beverage example emphasized the sensory and emotional aspect of the food in the story. The oil example used the 'slippery' property of food, where participants were encouraged to feel the 'slippery' oil in their mouth and then heard the impact of that 'slippery' oil play out in the story. Similarly, the 'calm' beverage used the affective / emotional property of the food, where participants drank something that they found calming, and heard its calming effect play out in the story. On the contrary, giving Gummy healthy food to build a boat did not leverage any properties of the food. Perhaps, if the physical properties of food were used such as foods that float gave gummy a boat, the consequence would be more believable.

2) The properties of the participant's actions of consuming the food had a logical narrative consequence on their narrative body and Gummy. The 'slippery' oil made the walls of the stomach slippery, which led to Gummy's fall. The 'calm' drink (and deep breaths), helped slow down the diegetic body's heart rate, giving Gummy an easier ride. Just asking the participant to perform an action that brings attention to their lived body is not enough. Without a logical consequence, the participant may ignore the how the action feels or even get distracted by it. For example, when Gummy called out how the belly rubs felt 'warm', some participants felt confused, even though they acknowledged the belly rubs did feel warm. This could have been because the warmth of those belly rubs had no narrative consequence. Having narrative consequences that leverage the sensory and emotional properties of the participant's actions (in this case, the food) can help participants buy into why their action led to a particular consequence, adding buy in, and giving them ontological roles.

Next, we describe how to leverage the sensory and emotional properties of food in an interactive narrative:

1) **Sensory**: Similar to Mueller et al.'s description of how a person's lived body can be engaged through localized sensations [27], we recommend bringing attention to one's body through the sensory properties of food such as the taste, temperature, smell, and texture. Different foods can feel juicy, sticky, hot/cold, oily, sugary, etc. We also recommend having a narrative consequence because of these sensory properties. Grounding this recommendation in a hypothetical example from GWO- juicy foods like an orange could have bathed the stomach walls and given Gummy a shower.

2) **Emotional**: Food can evoke emotional and nostalgic responses. Obrist et al. [30] described how to map taste to emotions, but we found that emotional responses to food can be very personal. To ensure that a participant actually feels the affective response, we recommend allowing them to personalize the interface– letting them interact with things that evoke the emotional response for them. In GWO, participants were asked to bring any drink that they found calming, and hence, they already associated their drink to a calm state when they experienced it in the story. The narrative consequence of such foods can be represented in the body through a narrative physiological response. For example, deep breaths and the calming drink in GWO made people feel calm and lower their heart rate. This concept of emphasizing the physiological response of an action has also been used in previous work like in Harley et al.'s work with the squirrel [17] and A Breathtaking Journey [24].

Lastly, we describe the above concepts further through a hypothetical example based on GWO. Let's assume that Gummy progressed in his journey when the participant drank cold milk. Here milk gives a tangible and ontological interface but does not tap into any of the soothing properties of milk. How is milk then much different from pressing a button? Building on this example further, what if drinking cold creamy milk made a soft and cool corner in the stomach for Gummy, where he could relax and go to sleep. This example takes the sensory (soft, cool) and emotional (relaxing, sleepy) qualities of milk and weaves them into the story such that it has a logical impact on the body (soft and cool corner) and Gummy (helps him relax and sleep), which may help participants feel the action viscerally, leveraging the experience of their lived body. It also increases buy-in for why milk helps Gummy compared to the first example.

In summary, leveraging the sensory and emotional properties of food and having logical narrative consequences can help participants feel their actions viscerally, engage their lived body, increase their buy in, and add more stakes in the story.

### 7.3 Leverage the diegetic body for character connection

The guidelines above to help participants feel that their bodies are a part of the story world. Beyond that, the following design recommendations can be used to leverage the diegetic body to create a sense of connection with the character.

The diegetic body can increase personal stakes for the participant in the narrative as their bodies are a part of the story, increasing the gravity of their choices. We recommend aligning the participant's personal stakes and goals with the character's goals in the narrative. In GWO, helping or hurting the body helped or hurt Gummy, largely aligning the goals of the participants and character. This goal alignment can help participants root for the character and feel for the character's success or failure. For example, participants mentioned feeling bad when the acid attack hurt Gummy and their stomach. They cared for Gummy, felt positive when they helped him, and were careful when they hurt Gummy.

Endowing participants with the character's well-being can also help them feel responsible for the character [13,36]. In GWO, they were endowed with Gummy's well-being who was dependent on their bodies' well-being. They felt responsible for Gummy since he was in their body – an entity they owned, cared for, and controlled. Furthermore, they could not directly impact Gummy but only help him on his journey. This put them more in a facilitator role and less in a God-like role, which may have fueled sympathy for Gummy. We recommend that participants can be endowed with the character's well-being by making the character dependent on an entity that participants have ownership, control, and responsibility for, in this case – the diegetic body.

Participants also felt responsible for Gummy because they chewed and hurt him in the beginning, which made them feel guilty for putting him in the situation he was. This act made participants pledge their allegiance to Gummy from the very beginning. We recommend creating hooks where participants may take actions that accidentally put the character in a difficult situation. These hooks can help participants feel responsible for helping the character.

## 8  CONCLUSION

Although there is a growing interest in designing and understanding playful interactions with food and technology, the intersection of food and interactive narratives has received much less attention. The few works that exist in this area often position people as passive observers. To further explore design possibilities in this area, we presented Gummy's Way Out (GWO) – a food based interactive narratives where participants eat a talking gummy bear and then help him find his way out of their bodies by eating various food items and performing different bodily actions. Besides giving participants the ability to be a part of the story world and impact the story (internal-ontological role), we explored how the 'diegetic' and 'lived' body added to the narrative experience. After conducting a study with 19 volunteers, we found that the diegetic and lived elements of the experience allowed participants to buy into the narrative consequences, added personal stakes in the story, increased the gravity of their choice, and made them feel more responsible for Gummy as he was in THEIR body. However, there were times when participants lost track that the story was happening in their bodies, and they had a hard time believing any anatomical inconsistencies in the narrative body. Based on the analysis of our findings, we give three design takeaways for the diegetic lived body in food based interactive narratives. 1) Create body tethers that remind people that THEIR body is in the story world. This can be done by asking them to perform abstract actions and being aware of the fragility of suspension of disbelief while designing with the body. 2) Leverage the emotional and sensorial aspects of food and the body – we recommend calling attention how the participant's action feels (emotionally or sensorily) in the story and then impacting the narrative body and the characters / story accordingly. This can help people buy into why their action led to a particular consequence giving them internal-ontological roles. It can also help people feel their actions viscerally, through which they can experience their lived bodies. 3) Leverage the diegetic body for character connection -

The diegetic body can increase the participant's personal stakes. Character connection can be strengthened if the participant's goals are aligned with the character's goals. We hope this work inspires more design opportunities for food and interactive narratives where participants take more active roles and engage with the sensory and emotional aspects of food and the body. We also hope this work provides insight into designing for the 'diegetic' and 'lived' body in tangible interactive narratives.